\documentclass[prb,a4paper,twocolumn,floatfix,preprintnumbers,amsmath,amssymb,superscriptaddress]{revtex4-2}

\usepackage{float}
\usepackage{graphicx}% Include figure files
\usepackage{dcolumn}% Align table columns on decimal point
\usepackage{bm}% bold math
\usepackage{color}
\usepackage{xspace}
\usepackage{amsmath}
\usepackage{textcomp} % Use with pdfLaTeX

\graphicspath{{./}{figure/}}
\usepackage[colorlinks,plainpages=false,linkcolor=blue,urlcolor=blue,citecolor=blue,pdfpagemode=UseNone,pdfstartview=FitBH]{hyperref}
\usepackage{dcolumn,graphicx,color,booktabs,microtype,afterpage}
\usepackage[charter,greekuppercase=italicized]{mathdesign}

\newcommand{\tcr}[1]{\textcolor{black}{#1}}

\begin{document}

\title{Fluctuating magnetism in Zn-doped averievite with well-separated kagome layers}
%\title{Fluctuating Ground State in Kagome Material Averievite through Spatial Decoupling}
%\title{Achieving a fluctuating ground state in the Kagome material Averievite through spatial decoupling}
%\title{Fluctuating ground state in the Kagome material Averievite through spatial decoupling}
%\title{Magnetism in insulating Kagome material Averievite }

\author{G.\ Simutis}
\email{gediminas.simutis@psi.ch}
\affiliation{Laboratoire de Physique des Solides, Paris-Saclay University and CNRS, France}
\affiliation{PSI Center for Neutron and Muon Sciences CNM, 5232 Villigen PSI,  Switzerland}
\affiliation{Department of Physics, Chalmers University of Technology, SE-41296 G\"{o}teborg, Sweden}

\author{L. Su\'{a}rez-Garc\'{\i}a}
\affiliation{PSI Center for Neutron and Muon Sciences CNM, 5232 Villigen PSI,  Switzerland}

\author{H.\ Zeroual}
\affiliation{Laboratoire de Physique des Solides, Paris-Saclay University and CNRS, France}

\author{I. Villa}
\affiliation{Laboratoire de Physique des Solides, Paris-Saclay University and CNRS, France}
\affiliation{PSI Center for Neutron and Muon Sciences CNM, 5232 Villigen PSI,  Switzerland}

\author{M. Georgopoulou}
\affiliation{Department of Chemistry, University College London, 20 Gordon Street, London WC1H 0AJ, United Kingdom}
\affiliation{Institut Laue-Langevin,71 avenue des Martyrs, CS 20156, 38042 Grenoble Cedex 9, France}

\author{D. Boldrin}
\affiliation{School of Physics and Astronomy, University of Glasgow, Glasgow, G12 8QQ}

\author{C. N.~Wang}
\affiliation{PSI Center for Neutron and Muon Sciences CNM, 5232 Villigen PSI,  Switzerland}

\author{C. Baines}
\affiliation{PSI Center for Neutron and Muon Sciences CNM, 5232 Villigen PSI,  Switzerland}

\author{T. Shiroka}
\affiliation{PSI Center for Neutron and Muon Sciences CNM, 5232 Villigen PSI,  Switzerland}
\affiliation{Laboratorium f\"ur Festk\"orperphysik, ETH Z\"urich, CH-8093 Z\"urich, Switzerland}

\author{R. Khasanov}
\affiliation{PSI Center for Neutron and Muon Sciences CNM, 5232 Villigen PSI,  Switzerland}

\author{H.~Luetkens}
\affiliation{PSI Center for Neutron and Muon Sciences CNM, 5232 Villigen PSI,  Switzerland}

\author{B. F{\aa}k}
\affiliation{Institut Laue-Langevin, 71 avenue des Martyrs, CS 20156, 38042 Grenoble Cedex 9, France}

\author{Y.~Sassa}
\affiliation{Department of Physics, Chalmers University of Technology, SE-41296 G\"{o}teborg, Sweden}
\affiliation{Department of Applied Physics, KTH Royal Institute of Technology, SE-106 91, Stockholm, Sweden}

\author{M.~Bartkowiak}
\affiliation{PSI Center for Neutron and Muon Sciences CNM, 5232 Villigen PSI,  Switzerland}

\author{A. S. Wills}
\affiliation{Department of Chemistry, University College London, 20 Gordon Street, London WC1H 0AJ, United Kingdom}

\author{E. Kermarrec}
\affiliation{Laboratoire de Physique des Solides, Paris-Saclay University and CNRS, France}

\author{F. Bert}
\affiliation{Laboratoire de Physique des Solides, Paris-Saclay University and CNRS, France}

\author{P. Mendels}
\affiliation{Laboratoire de Physique des Solides, Paris-Saclay University and CNRS, France}

\date{\today}

\begin{abstract}

Kagome lattice decorated with $S=1/2$ spins is one of the most discussed ways to realize a quantum spin liquid. However, all previous material realizations of this model have suffered from additional complications, ranging from additional interactions to impurity effects. Recently, a new quantum kagome system has been identified in the form of averievite Cu$_{5-x}$Zn$_x$V$_2$O$_{10}$(CsCl), featuring a unique double-layer spacing between the kagome planes. Using muon spin spectroscopy we show that only a complete substitution (i.e. $x=2$) of interplanar copper ions leads to a quantum-disordered ground state. In contrast, the parent compound ($x=0$) exhibits long-range magnetic order, with a phase transition around 24\,K. Experiments performed on the partially substituted material ($x=1$) show that the transformation proceeds through an intermediate disordered, partially frozen ground state, unaffected by pressures up to 23\,kbar. Our study provides a microscopic view of the magnetism of the decoupling of the kagome layers and establishes the averievite as a new material platform for the experimental study of the fully-decoupled kagome layers. \end{abstract}

\pacs{}
%\maketitle\enlargethispage{3pt}
\maketitle{}
\section{Introduction}
A quantum spin liquid is one of the most sought-after states of matter, in which interacting magnetic moments do not form long-range order, but remain fluctuating in a highly-entangled manner \cite{Anderson1973,Anderson1987,Balents2010,Savary2016,Broholm2020}. Geometrically frustrated lattices offer an important way to realize this state, with the kagome lattice being the most promising candidate among the nearest-neighbor models \cite{Hastings2000,Ran2007}.

While there is a general agreement that the antiferromagnetic kagome lattice hosts a quantum spin liquid state, its nature and topological classification are still under debate. One of the intriguing aspects is the stability of the spin-liquid phase. While it is expected to be a very fragile state, it nevertheless survives the addition of a small antisymmetric Dzyaloshinskii–Moriya (DM) interaction \cite{Cepas2008}, whose critical value is still under debate \cite{Lee2018}. Another extensively debated issue is the presence (or absence) of an excitation gap in the low-energy spectrum, with theoretical works suggesting both gapped~\cite{Jiang2008,Yan2011,Depenbrock2012} and gapless \cite{Ran2007,Iqbal2013,Iqbal2015} scenarios. Recent tensor-network results tend to suggest a gapless Dirac-like spectrum of spinons ~\cite{Liao2017}. In view of the theoretical challenges, an experimental input is essential for further progress.

Extensive experimental studies have been performed on several kagome candidates. The most prominent ones are herbertsmithite and barlowite, since they come closest to the simple nearest-neighbor Heisenberg antiferromagnetic model on the kagome lattice~\cite{Shores2005,Mendels2007,Tustain2020,Smaha2020}. These candidate materials show no long-range order, which is an essential first criterion for the quantum spin liquid. However, most of the information about that quantum state resides in its dynamical properties, which have proven to be much more difficult to pin down. An emerging conclusion from multiple experimental techniques is that
the various material realizations exhibit a combination of
a pure kagome behavior with a response from correlated
defects. Extensive NMR studies observed different spin-lattice relaxation rates, depending on the site under investigation, thus suggesting that the dynamics is strongly affected by the proximity of defects~\cite{Khuntia2020,Wang2021}. Similarly, neutron scattering experiments have revealed that the momentum dependence of the dynamical structure factor can change in the low-energy regime, if the defect contribution becomes dominant, making the evaluation of the pure kagome behavior increasingly complicated \cite{Han2012,Han2016,Punk2014,Breidenbach2025}.

\begin{figure} [b]
\centering
\includegraphics[width={\columnwidth}]{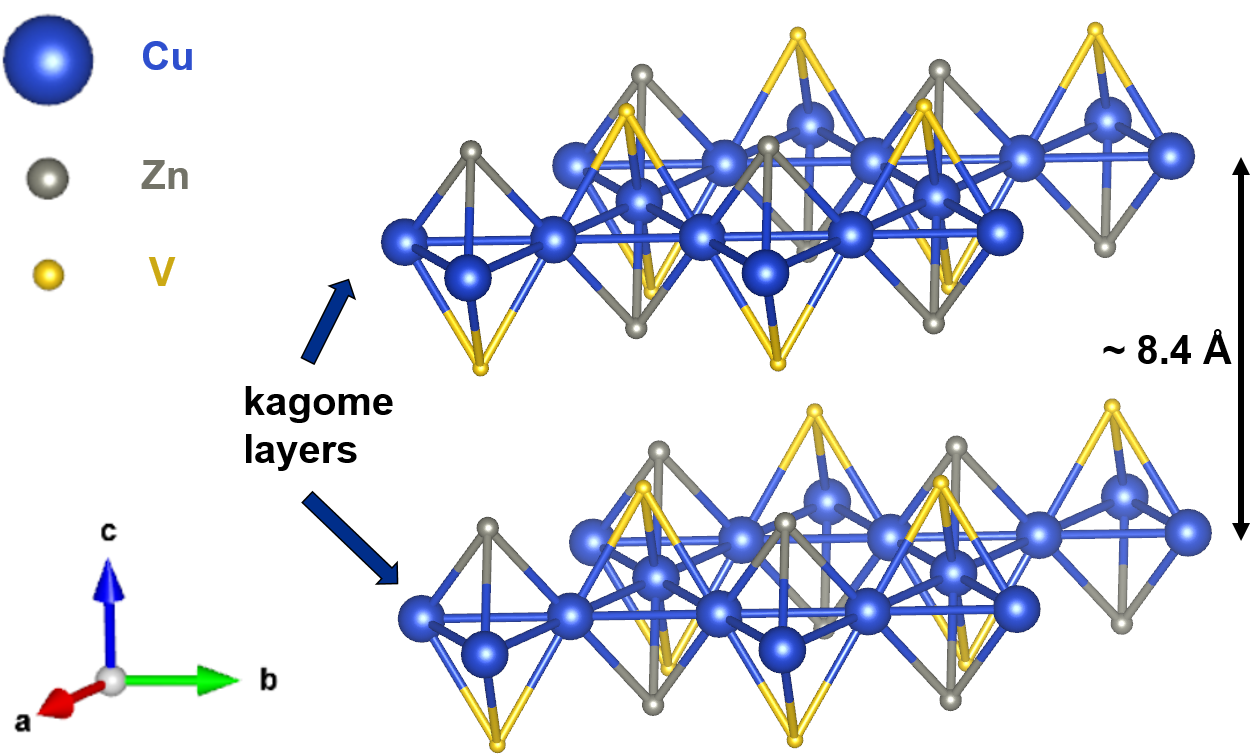}\\
\caption{\label{fig:structure}%
Crystal structure of the averievite compound, when all the interlayer Cu atoms are replaced by Zn. The kagome planes are built from corner-sharing triangles of copper atoms. The two interlayers host honeycomb lattices of Zn and V atoms. For clarity, the Cs, Cl and O atoms are omitted.
}
\end{figure}

The problem of identifying and separating the contribution of defects is especially pronounced, because even in  compounds of nominally depleted interplanar copper ions, only a single non-magnetic layer separates the two-dimensional kagome planes \cite{Shores2005,Tustain2020}. Clearly, even a minute number of inter-plane impurities can affect simultaneously two kagome planes, potentially in a coupled manner \cite{Han2016}. As such, it is imperative to investigate kagome materials with different arrangements and types of imperfections to shed light on the relevant contributions and to disentangle them.

In the present work we tackle this problem by studying a new quantum spin liquid candidate via the muon-spin rotation and relaxation ($\mu$SR) technique. Recently, it was realized that averievite, a material created by
the volcanic eruption of Tolbachik in Kamchatka~\cite{Vergasova1996},
can host kagome planes separated by two honeycomb layers. This structure is illustrated in Fig.~\ref{fig:structure} with its distinctly different distribution of the kagome layers compared to the existing compounds. In particular, the two honeycomb layers can be modified and gradually depleted of magnetic ions, because the copper sites are preferentially replaced when Zn is substituted for Cu \cite{Botana2018,Biesner2022,Liu2024}. The spacing between the two adjacent kagome layers is $\approx$ 8.4 \AA, which is significantly larger than, for example, that in herbertsmithite, where the layers are separated by $\approx$ 5.1 \AA~only \cite{Shores2005}. Further, the material could be synthesized in the lab up to a concentration of $x=1$ and bulk measurements of Cu$_{5-x}$Zn$_x$V$_2$O$_{10}$(CsCl) showed that the magnetic ordering transition is suppressed upon Zn substitution \cite{Botana2018}. Subsequently, the substitution level of Zn was increased all the way up to $x = 2$ \cite{Georgopoulou2023}, thus providing the opportunity to systematically study the magnetism of averievite throughout the full substitution range, up to fully decoupled kagome layers.

In our $\mu$SR experiments, we follow the evolution of the magnetism of averievite upon replacing Cu atoms with Zn in the synthetic version of Cu$_{5-x}$Zn$_x$V$_2$O$_{10}$(CsCl)~\cite{Botana2018,Georgopoulou2023}. This allowed us to extend the current studies and cover the full replacement range of Cu$^{2+}$ by Zn$^{2+}$ in the inter-kagome layers ($x = 0$, $1$, $2$). In case of an optimal decoupling of the adjacent kagome layers, we expect to access their purely two-dimensional physics. In the following, we refer to the compounds with different zinc content, \tcr{$x = 0$,} 1, and 2, as Zn0, Zn1, and Zn2, respectively.

We confirm the long-range order in the Zn0 parent compound and find that
the fully-substituted Zn2 system has a dynamically fluctuating ground state.
The partially substituted Zn1 case exhibits a suppression of the long-range order, as reported previously~\cite{Botana2018}. However, we find that at this substitution level, the electronic spins attain a complex state, characterized by a disordered and inhomogeneous behavior, including a partial freezing. This glassy phase eventually morphs into a dynamical state, dominated by fast magnetic fluctuations, when all the Cu interlayer sites are replaced by Zn ($x = 2$).

\section{Experiment}

The polycrystalline samples of Cu$_{5-x}$Zn$_x$V$_2$O$_{10}$(CsCl) were synthesized by a solid state reaction from the stoichiometric proportions of the CsCl, ZnO, CuO, and V$_2$O$_5$ powders \cite{Georgopoulou2023}. Muon-spin relaxation ($\mu$SR) measurements were performed using the continuous muon beam at the Paul Scherrer Institute, Villigen, Switzerland. To fully characterise the material, a combination of the GPS \cite{Amato2017}, Dolly and GPD \cite{Khasanov2016} spectrometers was used. Experiments at the GPS spectrometer were conducted by cooling the samples using a helium-flow cryostat to reach temperatures down to 1.6 K. Measurements down to 0.27 K at GPD and Dolly spectrometers were performed using a $^3$He Heliox Cryostat. For the high-pressure experiment, a two-walled low-background CuBe/MP35 cell was employed \cite{shermadini2017} with pressure transmitted to the sample via Daphne oil 7373. The pressure was determined by tracking the superconducting transition temperature of a small piece of Indium placed in the pressure cell. As the pressure-cell response at low temperatures may be slightly cell dependent, the exact same temperature points were measured at zero pressure and with applied pressure. The data analysis from all of the experiments was performed using the \textit{musrfit} program \cite{Suter2012}.
\begin{figure}
\centering
\includegraphics[width={\columnwidth}]{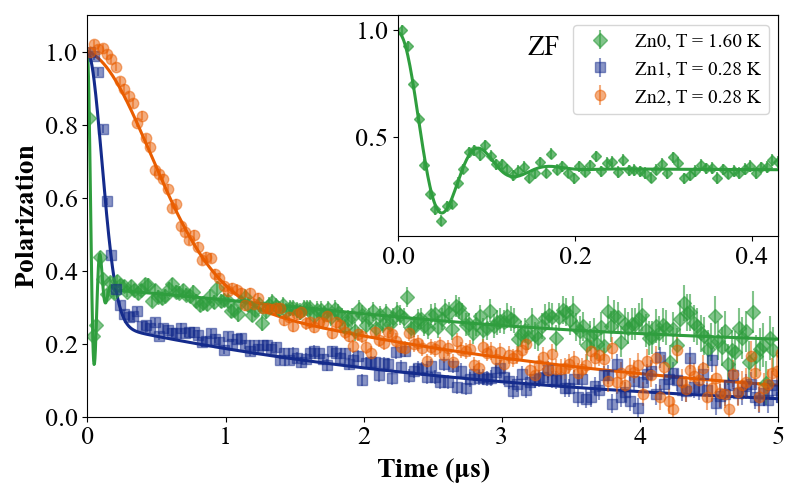}\\
\caption{\label{fig:muonZF}%
Muon spin polarization as a function of time, implanted into different versions of the averievite compound. The $x=0$ compound exhibits well-defined oscillations of the muon spin polarization, indicative of long-range order. By contrast, the Zn-substituted versions exhibit no oscillations, but the response is dependent on the substitution level. While $x=$1 exhibits a rapid depolarization, the fully-decoupled $x=2$ compound remains fully dynamic.
}
\end{figure}

\section{Results}

We first address the ground state properties of the system, measured at low temperatures (T = 0.27 K for Zn1 and Zn2 compounds and T = 1.6 K for the pure copper system) as the composition is varied. $\mu$SR  has an excellent sensitivity to small magnetic moments (down to 10$^{-4} \mu_B$) and the response of the muon spin is qualitatively different in the cases of long-, and short-range order as well as dynamically fluctuating ground states. Static, long-range ordered magnetic materials induce local fields at the muon implantation sites, which lead to a well-defined precession of the muon spin. This scenario is realized in the x=0 system as can be seen in Fig.~\ref{fig:muonZF}. At the opposite end of the solid solution, the muon response of the $x=2$ compound shows neither oscillations, nor a fast depolarization of the muon spin, which suggests that the system remains in a fluctuating regime. The intermediate substitution of x = 1 shows an even more complex behavior, without a precessing signal, but with a fast depolarization of the muon spins seen in Fig.~\ref{fig:muonZF} which arises from a combination of static and dynamic effects. In the following paragraphs, we detail the analysis that leads to such conclusions. 

\subsection{Parent compound with $x=0$}

We begin by addressing the results of the synthetic parent compound of averievite, where both the kagome and honeycomb layers are occupied by Cu$^{2+}$. Earlier 
measurements~\cite{Botana2018,Georgopoulou2023,Guchhait2024} have reported a magnetic transition at T$_N=$ 24 K. At base temperature, the muon spin asymmetry exhibits well-defined, yet damped oscillations. We have found that the muon asymmetry as a function of time is best reproduced by a sum of a Bessel function (the first term in \ref{eq:Zn0}, multiplied by a gaussian relaxation, with rate $\sigma$, arising from a distribution of the fields) and a slowly relaxing fraction, arising from the internal field contribution parallel to the initial muon spin polarization (also known as $1/3$-tail as this is a value expected in a randomly oriented polycrystalline samples). The need to add the second exponential relaxation term $\lambda_{tail}$ suggests that there are some persistent dynamics even deep in the ordered phase, which is often observed in ordered magnets that have frustrated motifs \cite{Somesh2021}.

At high temperatures, the data can be well reproduced by a Kubo-Toyabe function, with the asymmetry time evolution originating from nuclear spins, in the fast fluctuating limit for Cu$^{2+}$ electronic spins. In order to evaluate the temperature dependence of the muon spectrum and describe the behavior throughout the whole measured temperature range, we added a high-temperature response as the second term in \ref{eq:Zn0}, with $f_F$ representing the magnetically ordered fraction, indeed not described by the high-temperature model. The whole equation reads:

\begin{equation}
\begin{aligned}
\label{eq:Zn0}
P(t) = f_F \big[ (\frac{2}{3}) J_0(t) e^{\frac{-\sigma^2 t^2}{2}} + \frac{1}{3} e^{-\lambda_{tail}t}\big] + \\ (1 - f_F)P_{KT}(\Delta,t) e^{-\lambda_P t}.
\end{aligned}
\end{equation}

Equation \ref{eqn:w_Bes} describes the zeroth-order spherical Bessel function $J_0$, with $B_{max}$ corresponding to the maximum value of the field distribution. The response at high temperature - the Kubo Toyabe $P_{KT}$ is given by equation \ref{eq:GKT}.

\begin{equation}
\label{eqn:w_Bes}
\begin{aligned}
J_0 (t) = \frac{\sin(\gamma_\mu B_{max} t)}{\gamma_\mu B_{max} t} \mathrm{.}
\end{aligned}
\end{equation}

\begin{equation}
\begin{aligned}
\label{eq:GKT}
P_{KT}(t) = \frac{1}{3} + \frac{2}{3} (1 - \gamma_\mu^2 \Delta_G^2 t^2) \exp(-\frac{\gamma_\mu^2 \Delta_G^2 t^2}{2}) \mathrm{.}
\end{aligned}
\end{equation}

\begin{figure}
\centering
\includegraphics[width={\columnwidth}]{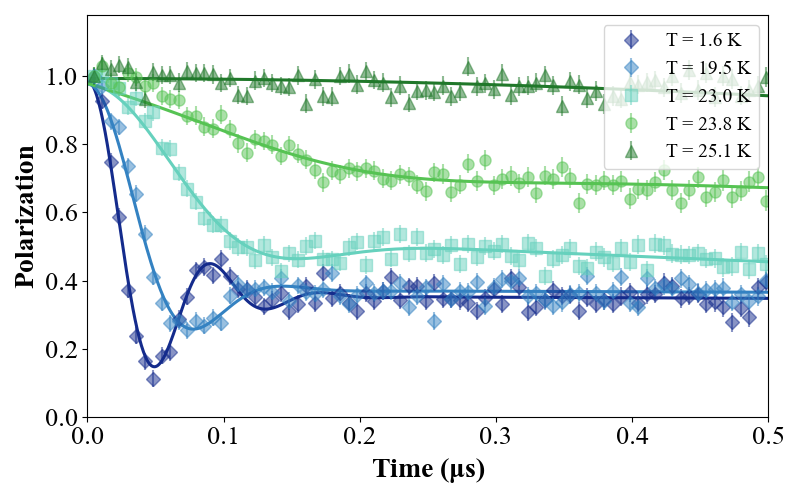}\\
\caption{\label{fig:ZF_Zn0_spectrum}%
Muon spin polarization as a function of time for several temperatures for the parent compound. The solid lines are fits to Eq.~\eqref{eq:Zn0}, where the Gauss Kubo-Toyabe function dominant at high temperatures develops into an oscillating function upon cooling down.
}
\end{figure}

\begin{figure}
\centering
\includegraphics[width={\columnwidth}]{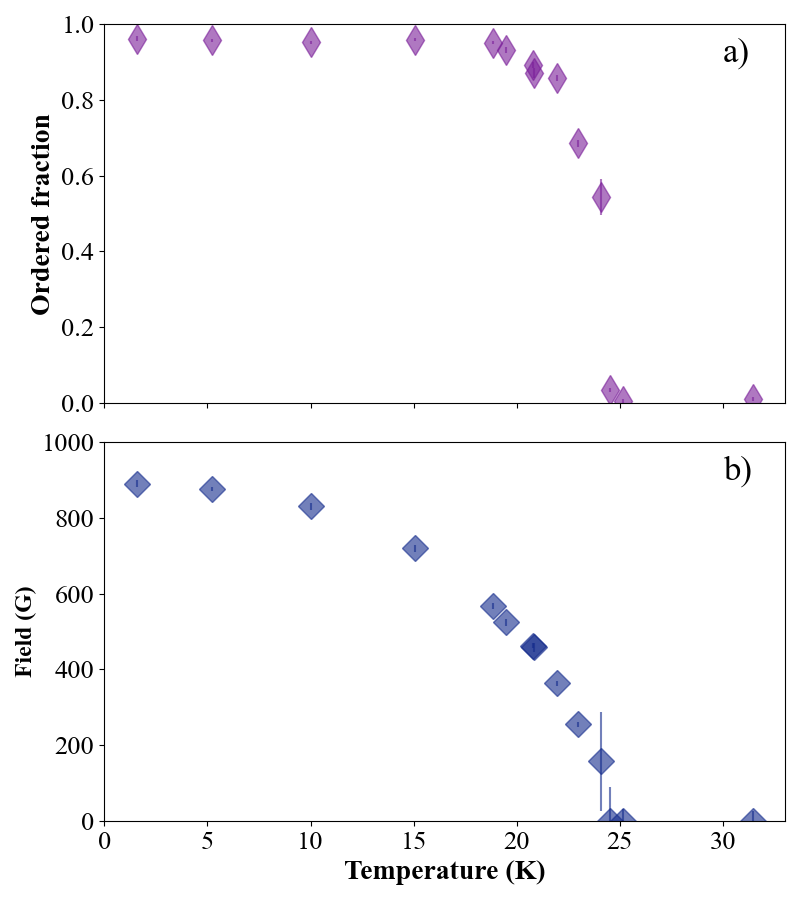}\\
\caption{\label{fig:ZF_Zn0_params}%
(a) Temperature dependence of the magnetically ordered fraction, obtained using Eq.~\eqref{eq:Zn0}. (b) Temperature dependence of the characteristic internal field $B_{max}$, obtained from the oscillatory parameter of the fit function from Eq.~\eqref{eqn:w_Bes}.
}
\end{figure}

Several spectra at different temperatures are presented in Fig.~\ref{fig:ZF_Zn0_spectrum}, together with the corresponding fits. Figure~\ref{fig:ZF_Zn0_params} (a) shows the magnetically ordered sample fraction, the onset of which is consistent to the previously reported transition temperature of 24 K. Figure~\ref{fig:ZF_Zn0_params} (b) shows the temperature dependence of the internal field, which is the order parameter of the magnetically ordered state.

\subsection{Towards decoupling of Kagome layers in $x=2$ compound}

\begin{figure}
\centering
\includegraphics[width={\columnwidth}]{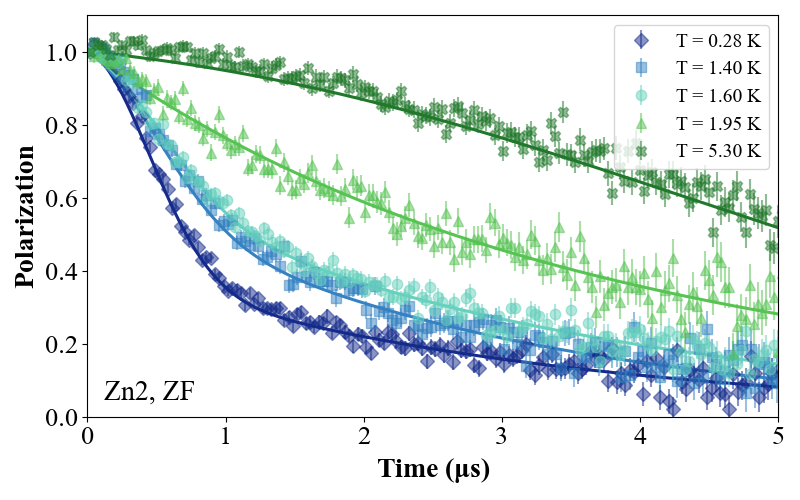}\\
\caption{\label{fig:ZF_Zn2_spectrum}%
Muon spin polarization as a function of time for several temperatures for the fully decoupled compound with $x=2$. The solid lines are fits to Eq.~\eqref{eq:Zn1_Zn2}.
}
\end{figure}

When all of the Cu$^{2+}$ ions in the interlayers are replaced by Zn$^{2+}$, the $\mu$SR response drastically changes. Neither oscillations are observed down to the lowest measured temperature of 270 mK, nor a 1/3 tail can be observed. The time evolution of the muon decay asymmetry at selected temperatures is displayed in Figure~\ref{fig:ZF_Zn2_spectrum}. At high temperatures, a slow depolarization is observed, which, as before, can be well reproduced by a Kubo-Toyabe function. At low temperatures, the depolarization rate increases and the rather complex response is best described by a sum of two functions - a faster gaussian relaxation that accounts for the early-time behavior and the slower tail-like relaxation. In order to obtain a best fit throughout the whole temperature range, a sum of an exponential relaxation and a fast gaussian relaxation is used, multiplied by a Kubo-Toyabe function $P_{KT}$:

\begin{equation}
\begin{aligned}
\label{eq:Zn1_Zn2}
A(t) = & A_0 P_{KT}(t) \big[ f e^{\left(-\frac{1}{2} (\sigma t)^2 \right)} + (1-f) e^{-\lambda t} \big],
\end{aligned}
\end{equation}

\begin{figure}
\centering
\includegraphics[width={\columnwidth}]{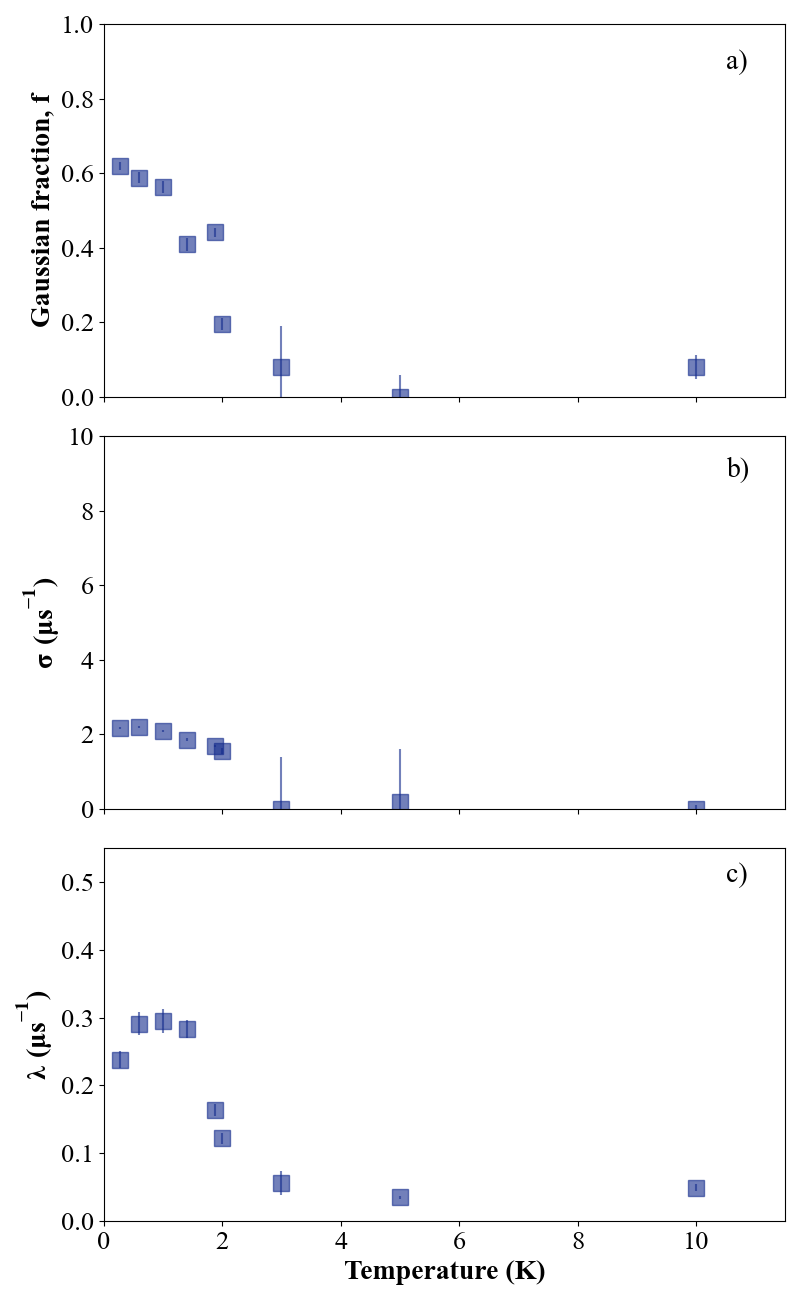}\\
\caption{\label{fig:ZF_Zn2_params}%
Extracted parameters of the Zn2 system at various temperatures, obtained via Eq.~\eqref{eq:Zn1_Zn2}. (a) Temperature dependence of the fraction of the signal that shows Gaussian depolarization. (b) Temperature dependence of the Gaussian depolarization rate, as obtained from Eq.~\eqref{eq:Zn1_Zn2}. (c) Temperature dependence of the relaxation rate, originating from magnetic fluctuations. 
}
\end{figure}

The relaxation of static nuclear origin associated with the KT can be fixed at the high temperature value with the reasonable assumption of $f=0$ and fast electronic spins dynamics ($\lambda = 0$). We observe that both the gaussian relaxation and the exponential relaxation increase below 2 K, with the gaussian function dominating the low-temperature response, see Fig.~\ref{fig:ZF_Zn2_params}. Similar increases of the depolarization rates at low temperatures have been observed in multiple quantum spin liquid candidates~\cite{Mendels2007, Kermarrec2011,Clark2013,Li2016,Majumder2020}.

\begin{figure}
\centering
\includegraphics[width={\columnwidth}]{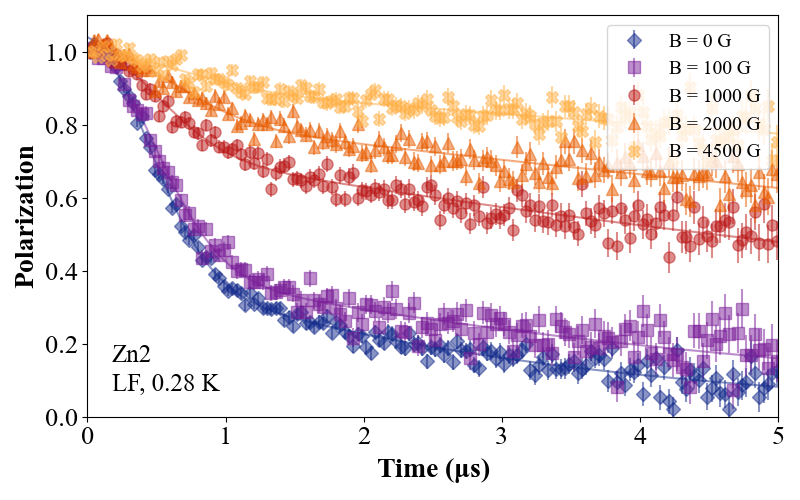}\\
\caption{\label{fig:LF_Zn2_spectrum}%
Muon spin polarization as a function of time for several longitudinal fields for the compound with $x=2$. Despite the slow relaxation rate observed in the zero-field experiment, substantial fields are needed to even partially decouple it, pointing to a dynamic ground state.
}
\end{figure}

In order to further investigate the origin of the muon depolarization, we performed longitudinal field measurements, which are displayed in Figure~\ref{fig:LF_Zn2_spectrum}. If the ground state were static, the gaussian relaxation component $\approx$ 2 $\mu s^{-1}$, would correspond to an internal field distribution of width $\Delta$B $= \sigma / \gamma_\mu$ $\approx$ 24 G. One would then expect a substantial decoupling in a 100 G field which is not the case. Moreover, the exponential relaxation persists upon application of even larger longitudinal fields. All this confirms the dynamical origin of the relaxation and points to persistent spin fluctuations, with a relaxation plateau at low temperature.

\subsection{Partially substituted compound with $x=1$}

\begin{figure}
\centering
\includegraphics[width={\columnwidth}]{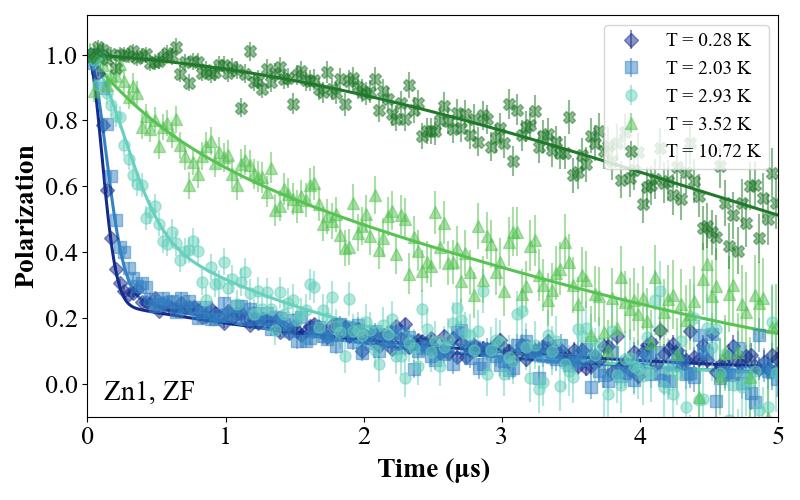}\\
\caption{\label{fig:ZF_Zn1_spectrum}%
Muon spin polarization as a function of time for several temperatures for the partially substituted compound with $x=1$. At high temperatures the response arises from randomly-oriented nuclear magnetic moments and electronic relaxation emerging at low temperatures with the steep component originating from the frozen spins as described in text. The solid lines are fits to Eq.~\eqref{eq:Zn1_Zn2}.
}
\end{figure}

When half of the Cu$^{2+}$ ions in the interlayers is replaced by Zn$^{2+}$, the $\mu$SR response shows some qualitative similarities to the fully substituted Zn2 case, with the time evolution of the muon decay asymmetry at selected temperatures displayed in Figure~\ref{fig:ZF_Zn1_spectrum}. At high temperatures, a slow gaussian depolarization is observed, which as before, can be well reproduced by a Kubo-Toyabe function, arising from nuclear magnetic moments. At low temperatures the depolarization rate increases and the rather complex response is again best described by the sum of two functions - a fast gaussian relaxation that accounts for the early-time behavior and the slow tail-like relaxation. No oscillations are observed down to the lowest measured temperature of 270 mK. 

\begin{figure}
\centering
\includegraphics[width={\columnwidth}]{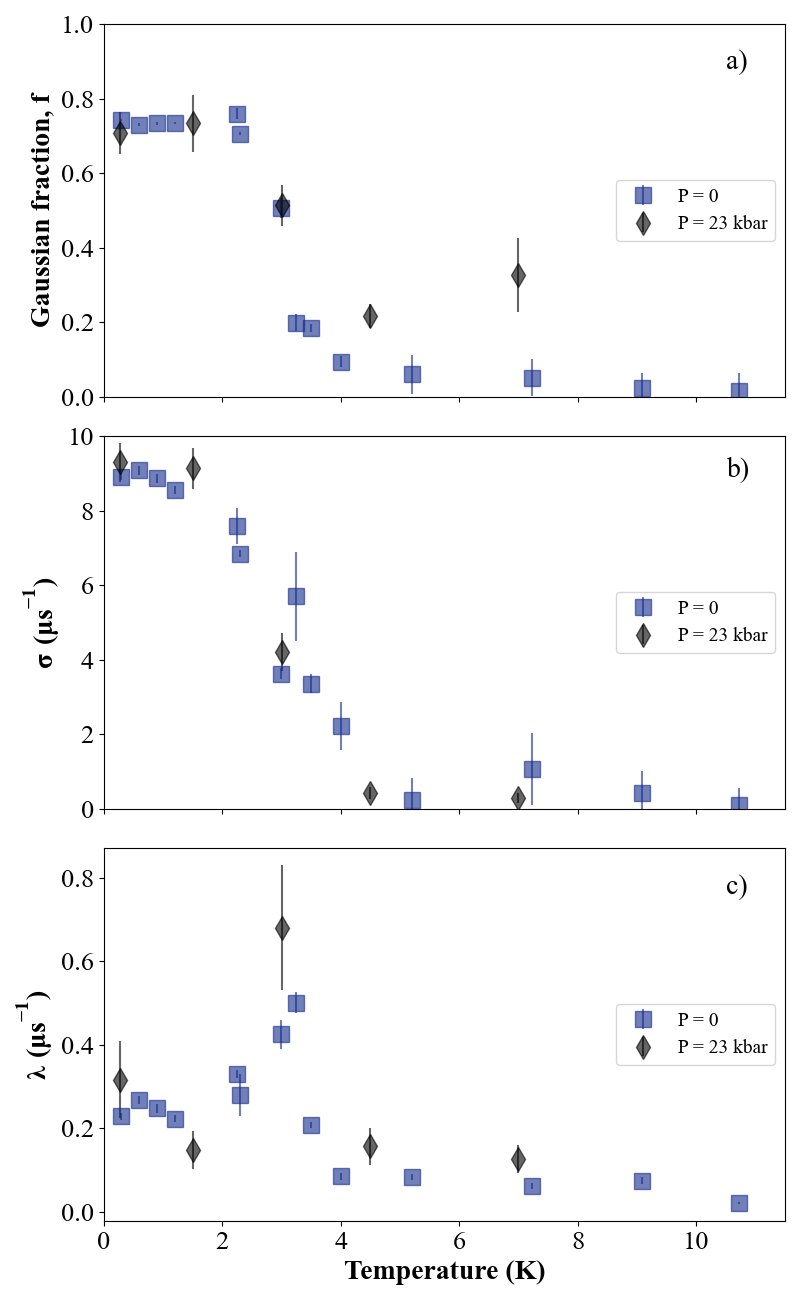}\\
\caption{\label{fig:ZF_Zn1_params}%
The extracted parameters of the Zn1 system at various temperatures, measured at zero pressure and with the pressure applied. The values are obtained by fitting Eq.~\eqref{eq:Zn1_Zn2}. (a) Temperature dependence of the fraction of the signal that shows Gaussian depolarization. (b) Temperature dependence of the Gaussian depolarization rate, which corresponds to the width of the field distribution, as obtained from Eq.~\eqref{eq:Zn1_Zn2}. (c) Temperature dependence of the relaxation rate, originating from magnetic fluctuations.}
\end{figure}

However, the gaussian depolarization rate is four times faster than in the fully substituted case, suggesting a possibility of spin freezing. In order to understand this low-temperature phase and how it is established, we again study the behavior as a function of temperature, as before fitted by an empirical function of Eq.~\eqref{eq:Zn1_Zn2}, with the results displayed in Figure~\ref{fig:ZF_Zn1_params}. On one hand, we observe that the onset of the gaussian depolarization rate takes place at a higher temperature of 4 K. Additionally, we find that the exponential relaxation of the tail has a peak at around the same temperature before settling at a persistent low-temperature value of $\approx$ 0.3 $\mu s^{-1}$. Both observations may be pointing to a large distribution of fields at the muon site that could either be of static or dynamic nature.

\begin{figure}
\centering
\includegraphics[width={\columnwidth}]{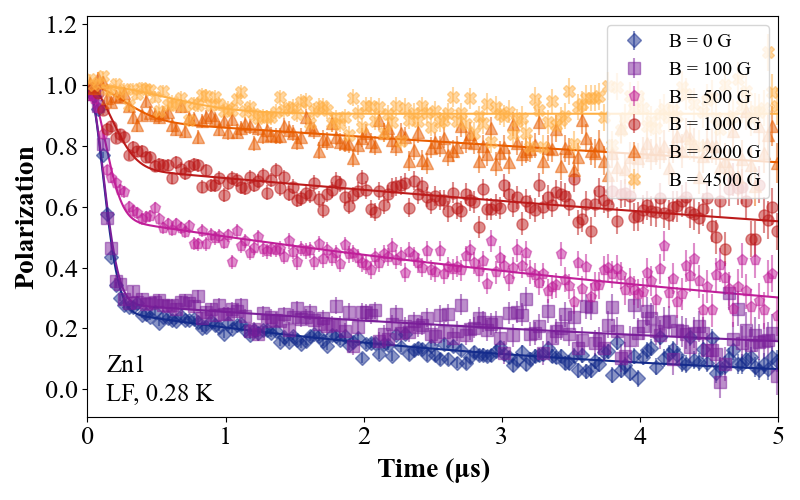}\\
\caption{\label{fig:LF_Zn1_spectrum}%
Muon spin polarization as a function of time for several applied longitudinal fields for the partially substituted compound with $x=1$. The solid lines are fits to Eq.~\eqref{eq:Zn1_Zn2}. Upon application of moderate fields, the fast gaussian relaxation is suppressed, but the slow relaxation persists.
}
\end{figure}

To further understand the nature of this complex phase, we performed longitudinal field measurements, with a few selected fields shown in Figure~\ref{fig:LF_Zn1_spectrum}. The measurements reveal that the fast gaussian depolarization is quickly suppressed with field. Nevertheless, the slow exponential depolarization persists. The gaussian relaxation rate saturates in zero field at a value of $\approx$ 9 $\mu s^{-1}$, which would correspond to an internal field distribution of width $\Delta$B = $\sigma / \gamma_\mu$ $\approx 106$ G in a static scenario. Longitudinal fields of the order of 1000 G $\sim$ 10 $\Delta$ B decouple a large fraction of the gaussian depolarization, however the decoupling remains incomplete, pointing to a quasistatic phase with strong persistent dynamics. Interestingly, at base temperature the fraction of the fast gaussian component $f$ is fairly close to 2/3, which in the static picture would indicate that the whole sample undergoes a transition to a frozen phase with a large distribution of fields at the muon site. Yet, the absence of a dip in the plot of the asymmetry versus time clearly demonstrates that it does not correspond to a spin glass freezing. At the same time, the exponential relaxation remains finite at the lowest temperatures, which in the static case would correspond to a local field of 3 G. As even fields of a few kG do not decouple this relaxation, it originates from the remanent slow dynamical fluctuation of electronic spins.

One possibility to explain the rapid relaxation could be due to a homogeneous disorder within a disordered static system with a collection of Gaussian field distributions. Such a scenario would result in a Gaussian-broadened GKT function, with no dip in the depolarization measurement \cite{Noakes1997,Alexanian2025}. However, we find that the expected dynamic response in the longitudinal field doesn't follow the predicted behavior as seen in Appendix A, suggesting that the origin of the fast depolarization arises from a more complex scenario.

Another scenario to explain a gaussian relaxation that is not decoupled by longitudinal fields has been proposed to arise if a dynamic state of singlet spin pairs coexists with a small number of unpaired mobile spins \cite{Uemura1994}. In such a case, for example originally seen in a kagome compound SrCr$_8$Ga$_4$0$_{19}$, the dynamic fluctuations of spins persist even at lowest measured temperatures despite an apparent spin-glass freezing. However, in such cases, a one-component relaxation is expected that fully depolarizes the muon spins \cite{Uemura1994}. This is in contrast with our observation of a two-timescales depolarization behavior with a substantial slow-relaxing tail, hence pointing to a different behavior, likely arising from partially frozen inhomogeneous magnetic clusters with strong persistent spin dynamics

Additionally, we study the possibility that the origin of the fast Gaussian relaxation is due to a proximity to a quantum phase transition. Such a fast depolarization rate could also indicate that the system is close to a long-range magnetic order. In such case a small external pressure might force the system to order and result in a $\mu$SR spectrum with oscillations. Several frustrated magnets exhibit strong modifications of muon response even at moderate pressures \cite{Kermarrec2017, Majumder2018, Chatterjee2025}. Our measurements at 23 kbar, however, have shown the same spectrum as in the non-pressurized case, indicating that the lack of magnetic order likely comes from the decoupling of the layers, with remaining frozen spins leading to the fast depolarization, as elaborated in the Discussion section. Moreover, the extracted muon spin depolarization parameters in the pressurized sample as a function of temperature exhibit the same temperature dependence, even though there are fewer data points, as seen in Figure \ref{fig:ZF_Zn1_params}.

\section{Discussion}

The question of how a highly-correlated collective spin state can be achieved without breaking the time-reversal symmetry has been key to understand the formation of quantum spin liquids and how they would manifest in real materials. Here we investigated the gradual suppression of magnetism in a recently identified kagome material averievite.

In the copper-only compound Cu$_5$V$_2$O$_{10}$(CsCl), a three-dimensional lattice of magnetic ions leads to a magnetically frozen state. Our measurements have confirmed the previously identified onset of magnetic order at 24 K \cite{Botana2018} and the observed oscillations in muon decay asymmetry established that the state is long-range ordered. This completely agrees with a recent neutron study where the magnetic Bragg peaks are observed below the ordering temperature \cite{Georgopoulou2023}. A scheme of interactions has been proposed from a study of spin wave dispersion, the complexity of which can be related to the low symmetry of the parent compound in a similar way as the parent compounds of quantum spin liquid candidates such as clinoatacamite or barlowite \cite{Norman2016,Han2014}.

It has been found that the structure has a higher symmetry when Zn progressively replaces Cu in the intermediate layers, leading to a perfect kagome geometry for the interactions for $x=1$ \cite{Botana2018} and $x=2$ \cite{Georgopoulou2023} cases. Potentially, this lower symmetry of the parent compound may result in persistent low-symmetry clusters even in the $x=1$ case. Such clusters could in turn be responsible for the rather unusual depolarization behavior
in the half-substituted compound  Cu$_4$Zn$_1$V$_2$O$_{10}$(CsCl). Indeed, the early \cite{Botana2018} and subsequent \cite{Georgopoulou2023} studies of magnetic properties reported a kink in susceptibility for the $x=1$ system, with no trace of long-range magnetic order observed with neutrons. The anomaly is found at around the same temperature $\sim 3.5$~K at which we detect a partial freezing with a large distribution of fields. This likely corresponds to short-range-correlated spins freezing in a frustrated arrangement, yet at a much lower temperature than in the antiferromagnetic $x=0$ phase and also at a much lower temperature than the magnetic interaction energy scale as expected in a highly frustrated antiferromagnet (frustration ratio $T_g/J \sim 1/30$). This observation, together with the presence of persistent dynamics, points to a picture very different from a mean field spin-glass.

Our high-pressure measurement also strongly limits the possibility of the system being at the verge of ordering. The fact that on average half of the interlayer sites are filled with magnetic coppers leading to such a disordered state makes a close resemblance to herbertsmithite where the switch from an antiferromagnetic phase to a quantum spin liquid occurs when Zn has replaced 50-60 \% coppers of the interlayer \cite{Mendels2007}. Similarly, Zn-barlowite has also been found to display a frozen-spin behavior at intermediate substitution levels \cite{Smaha2020, Tustain2020, Yuan2022}.

Finally, in the case of more decoupled layers in the Cu$_3$Zn$_2$V$_2$O$_{10}$(CsCl) compound we observe the absence of freezing with clear muon signatures in line with previous studies in the emblematic herbertsmithite and Zn-barlowite compounds. Much alike in the latter, the presence of a Curie tail in the susceptibility might still signal the presence of some amount of Cu's in the interlayer site which might act as presently not understood defects in the kagome physics and could impact the nature of the spin liquid state \cite{Khuntia2020,Wang2021} and possibly be at the origin of the reminiscent muon spin relaxation \cite{Kermarrec2011}. Whether such a family with the possibility of varying the interlayer composition offers the opportunity to better tune and/or understand such defects and reveal the corresponding physics remains an open issue and presents an avenue for future work.

Combining the above observations, the following picture of magnetism in the averievite family of compounds emerges. In the original compound with full copper occupation, long-range magnetic order is established, with minor persistent fluctuations of magnetic moments. Upon partial substitution of inter-layer coppers in the Zn1 compound, a complex phase is established, possibly consisting of a collection of partially frozen inhomogeneous magnetic clusters with strong persistent spin dynamics. Finally, upon the full double-layer substitution, in the Zn2 compound, the spins that were frozen, become fluctuating, signaling a non-trivial quantum state, consistent with a spin liquid phase.

Examining different families of kagome compounds, obtaining a perfect system with strictly equal nearest-neighbour interactions, no anisotropy and absence of disorder may be an unachievable goal. At the same time, having different types of impurities or imperfections across different families is important to separate intrinsic kagome physics from -potentially still interesting - tangential effects. In this respect, averievite material family offers a new viewpoint due to the two spacer-layers between the kagome planes.

To summarize, we have performed the first microscopic study of magnetism in a quantum kagome lattice compound with two-interlayer separation. We have discovered that upon substitution of Zn, the ground state gradually develops into a highly-correlated quantum paramagnet with all of the spins slowly fluctuating. Future theoretical and experimental investigation of this system will be enlightening, especially for determining the role of deviations from the ideal Heisenberg model in the ground state selection.

\begin{acknowledgments}
This project was supported by Agence Nationale de la Recherche under LINK (ANR-18-CE30-0022) project, the Swiss National Science Foundation Mobility grant P2EZP2-178604, PALM LabEx grant ANR-10-LABX-0039-PALM and European Union’s Horizon 2020 research and innovation program under the Marie Skłodowska-Curie grant agreement No 884104 (PSI-FELLOW-III-3i).

\end{acknowledgments}

\appendix

\section{Appendix A}

\begin{figure}
\centering
\includegraphics[width={\columnwidth}]{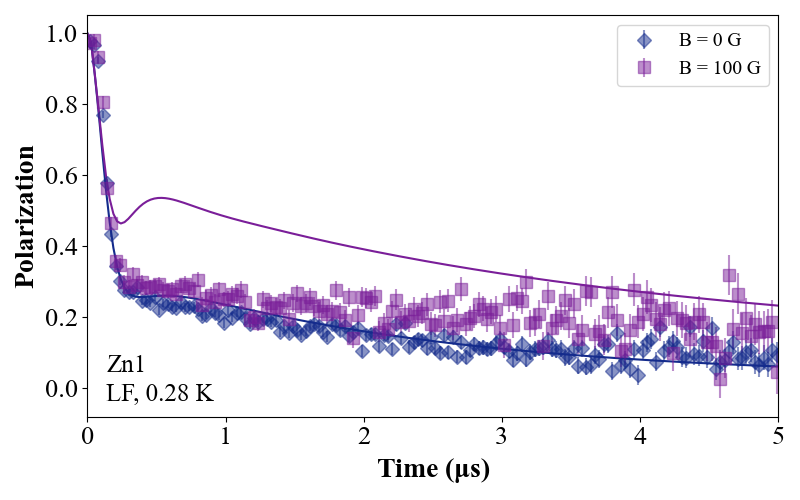}\\
\caption{\label{fig:LF_Zn1_GbGKT}%
Muon spin polarization as a function of time for zero field and an applied longitudinal field of 100 G for the partially substituted compound with x = 1. The solid lines are the expectations based on the gaussian-broadened GKT polarization function.
}
\end{figure}

Here we show the expected longitudinal field behavior for the Zn1 system, if the relaxation were described by a dynamic gaussian-broadened GKT function. As seen in Figure~\ref{fig:LF_Zn1_GbGKT}, the observed muon spin depolarization is not captured by this model, a more complex explanation for the ground state of the frozen spins at the intermediate substitution level is needed.

\bibliography{avebib}

\end{document}

\title{Supplementary information to MuSR and NMR study of kagome compound Averievite}

\author{G.\ Simutis}
\email{gediminas.simutis@psi.ch}
\affiliation{Laboratoire de Physique des Solides, Paris-Saclay University and CNRS, France}
\affiliation{Laboratory for Neutron and Muon Instrumentation,
Paul Scherrer Institut, CH-5232 Villigen PSI, Switzerland}
\affiliation{Department of Physics, Chalmers University of Technology, SE-41296 G\"{o}teborg, Sweden}

\author{M. Georgopoulou}
\affiliation{Department of Chemistry, University College London, 20 Gordon Street, London WC1H 0AJ, United Kingdom}
\affiliation{Institut Laue-Langevin,71 avenue des Martyrs, CS 20156, 38042 Grenoble Cedex 9, France}

\author{I. Villa}
\affiliation{Laboratoire de Physique des Solides, Paris-Saclay University and CNRS, France}

\author{D. Boldrin}
\affiliation{School of Physics and Astronomy, University of Glasgow, Glasgow, G12 8QQ}

\author{B. Fak}
\affiliation{Institut Laue-Langevin,71 avenue des Martyrs, CS 20156, 38042 Grenoble Cedex 9, France}

\author{A. S. Wills}
\affiliation{Department of Chemistry, University College London, 20 Gordon Street, London WC1H 0AJ, United Kingdom}

\author{F. Bert}
\affiliation{Laboratoire de Physique des Solides, Paris-Saclay University and CNRS, France}

\author{P. Mendels}
\affiliation{Laboratoire de Physique des Solides, Paris-Saclay University and CNRS, France}

\date{\today}

\begin{abstract}
Additional information is given to support the main text. We show the detailed muSR results, compare the spectra obtained from the Zn1 compound and display the subtraction procedure for the Zn2 system.
 \end{abstract}

\pacs{}
%\maketitle\enlargethispage{3pt}
\maketitle{}

\section{MuSR analysis and detailed results}

\begin{figure}
\centering
\includegraphics[width={\columnwidth}]{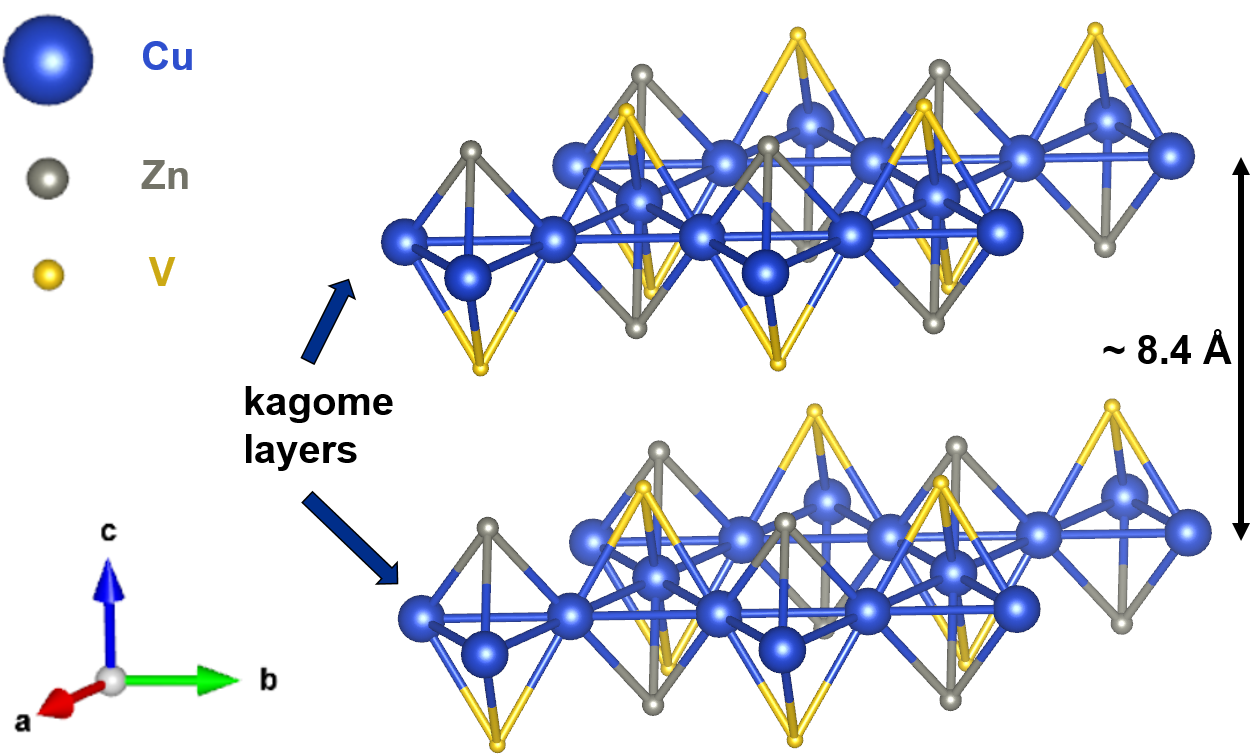}\\
\caption{\label{fig:structure}%
The left
}
\end{figure}

\begin{figure}
\centering
\includegraphics[width={\columnwidth}]{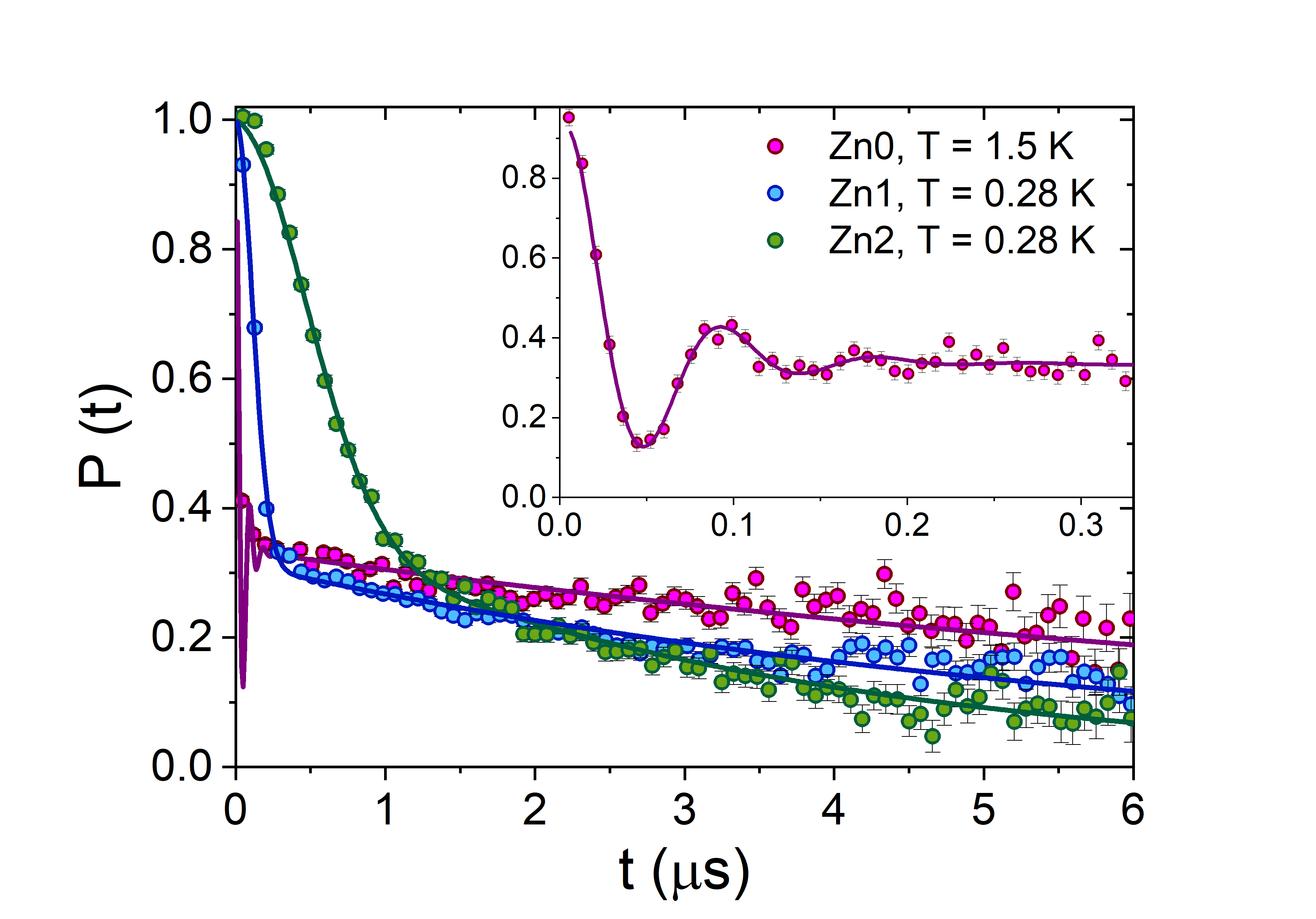}\\
\caption{\label{fig:muonZF}%
The left
}
\end{figure}

\begin{equation}\label{FF}
A (\bm{q})= \sum_i A_i exp(-i \bm{q r_i}) \mathrm{,}
\end{equation}

\bibliography{avebib}